\documentclass[aps,pre,twocolumn,superscriptaddress,10pt,reprint]{revtex4-2}
\usepackage[T1]{fontenc}
\usepackage[latin9]{inputenc}

\usepackage[unicode=true,pdfusetitle,
 bookmarks=true,bookmarksnumbered=false,bookmarksopen=false,
 breaklinks=false,pdfborder={0 0 0},pdfborderstyle={},backref=false,colorlinks=true]
 {hyperref}
 \hypersetup{
 citecolor=blue,
 urlcolor=blue,
 linkcolor=blue}
\usepackage[table,svgnames]{xcolor}
\usepackage{array}
\usepackage{units}
\usepackage{amsmath}
\usepackage{amssymb}
\usepackage{stackrel}
\usepackage{graphicx}
\usepackage{babel}
\usepackage{microtype}
\usepackage[final]{pdfpages}

\makeatletter
\AtBeginDocument{\let\LS@rot\@undefined}
\makeatother

\begin{document}

\title{Evidence of scale-free clusters of vegetation in tropical rainforests}

\author{Pablo Villegas}
\email[Corresponding author: pablo.villegas@cref.it]{}
\affiliation{`Enrico Fermi' Research Center (CREF), Via Panisperna 89A, 00184 - Rome, Italy.}
\affiliation{Instituto Carlos I de F\'isica Te\'orica y Computacional, Universidad de Granada, E-18071, Granada, Spain.}
\author{Tommaso Gili}
\affiliation{Networks Unit, IMT Scuola Alti Studi Lucca, Piazza San Francesco 15, 55100 - Lucca, Italy.}
\author{Guido Caldarelli}
\affiliation{DMSN, Ca' Foscari University of Venice, Via Torino 155, 30172 - Venice, Italy.}
\affiliation{European Centre for Living Technology (ECLT), Dorsoduro 3911, 30123 - Venice, Italy.}
\affiliation{Institute for Complex Systems (ISC), CNR, UoS Sapienza, Piazzale Aldo Moro 2, 00185 - Rome, Italy.}
\affiliation{London Institute for Mathematical Sciences (LIMS), W1K2XF London, United Kingdom.}
\author{Andrea Gabrielli}
\affiliation{`Enrico Fermi' Research Center (CREF), Via Panisperna 89A, 00184 - Rome, Italy.}
\affiliation{Dipartimento di Ingegneria Civile, Informatica e delle Tecnologie Aeronautiche, Universit\`a degli Studi 'Roma Tre', Via Vito Volterra 62, 00146 - Rome, Italy.}
\affiliation{Institute for Complex Systems (ISC), CNR, UoS Sapienza, Piazzale Aldo Moro 2, 00185 - Rome, Italy.}

\begin{abstract}
  Tropical rainforests exhibit a rich repertoire of spatial patterns emerging from the intricate relationship between the microscopic interaction between species. In particular, the distribution of vegetation clusters can shed much light on the underlying process that regulates the ecosystem. Analyzing the distribution of vegetation clusters at different resolution scales, we show the first robust evidence of scale-invariant clusters of vegetation, suggesting the coexistence of multiple intertwined scales in the collective dynamics of tropical rainforests. We use field data and computational simulations to confirm our hypothesis, proposing a predictor that could be particularly interesting to monitor the ecological resilience of the world's 'green lungs.'
\end{abstract}
\maketitle

A tantalizing hypothesis states that some biological systems may operate in the vicinity of a phase transition, fostering many functional advantages and optimizing the ability to react collectively \cite{MAM-RMP}. In light of such criticality hypothesis, there have been significant developments in understanding many real examples of inanimate natural phenomena as, for instance, sandpiles \cite{BTW}, earthquakes \cite{EarthQ}, or forest fires \cite{forest}. Subsequently, the advent of high-throughput technologies has led to found empirical evidence in living matter: from bacterial communities \cite{larkin2018} to the human heart \cite{kiyono2005}, networks of living neurons \cite{haimovici2013brain}, cluster of ants colonies \cite{vandermeer2008clusters} or gene expression \cite{nykter2008} (see \cite{MAM-RMP} for further examples).

All these systems share the presence of power-law distributed events, considered the hallmark of operating at (or close to) a second-order phase transition \cite{Bak}. For instance, \emph{neuronal avalanches}, i.e., cascades of activations
clustered in time, have been crucial to scrutinize the emergent dynamical behavior in neural populations \cite{BP2003}. However, unlike the von Neumann neighborhood in discrete systems, clustering statistics in continuous embeddings \cite{Villegas2022}, either temporal or spatial, rely on nearest-neighbor distance assumptions with an intrinsic degree of freedom: there is no unique way to define clusters in the system. The time-bin issue in determining neuronal avalanches is an example of this \cite{levina2017, MAM-RMP}.

Vegetation patterns are ubiquitous in arid and semiarid ecosystems \cite{rietkerk2008}. There, different large-scale patterns exhibit unambiguous spatial scales (e.g., Namibian fairy circles \cite{tarnita2017}), leading to long-lasting active debates and theoretical approaches \cite{kefi2007,rietkerk2002,rietkerk2008,martinez2014}. On the contrary, deciphering collective phenomena in large-scale saturated complex ecological systems such as rainforests, particularly the ecological patterns they show, represents a fundamental open problem in theoretical ecology \cite{Levin1992, Legendre1989, Sole2012}. Only a few studies have evidenced scale-free patterns, such as the low canopy gaps (after recent recolonizations) of Barro Colorado Island (BCI) \cite{sole1995} or different arid landscapes of the Kalahari, due to interacting effects of global resource constraints \cite{Scanlon}. Regarding semiarid environments, recent theoretical insights have demonstrated how environmental temporal variability can promote the emergence of vegetation patches with broadly distributed cluster sizes \cite{Paula}. Otherwise, cluster-based approaches have allowed, e.g., to identify the scales of spatial aggregation and the corresponding tree clusters in Malaysian tree species \cite{Plotkin}. However, how to cluster actual multivariate point patterns (beyond naive Newmann clustering tentatives) or, at least, how to extract their characteristic spatial scales, especially in saturated environments, is still an important issue to be solved.

Inferring properties of spatial point processes have been revealed to be essential for testing spatially related ecological theories and hypotheses \cite{law2009}. In particular, their analysis aims to explain the nature of underlying processes generating them and identify the scale at which they operate \cite{MoloneyPP}. However, different pieces must still be put back together in tropical forests: From dry deciduous forests to evergreen wet forests, empirical evidence suggests that most species are more aggregated than random \cite{Condit2000}, while anomalous density fluctuations, a fingerprint of long correlations, emerge in the spatial distribution of different tropical species \cite{VillegasRS}. Based on the dependence of species richness on the sampling area, different works suggested the possibility that specific ecosystems are in a state of incipient criticality \cite{Zillio2008,Volkov2004,Pascual2005}, even in the absence of power laws: the missing link to state that ecosystems can be placed in the vicinity of a critical point.

Through continuum clustering techniques \cite{Villegas2022}, here, we present evidence of scale-free vegetation clusters in empirical data from BCI, which sheds much light on their spatial aggregation properties and correlation scales. We compare current data with different simulated emergent spatial point processes, showing the wide variety of scales that play a crucial role in natural rainforests and uncovering emergent critical dynamics hitherto unknown. 

\begin{figure}[hbtp]
\begin{centering}
\includegraphics[width=1.0\columnwidth]{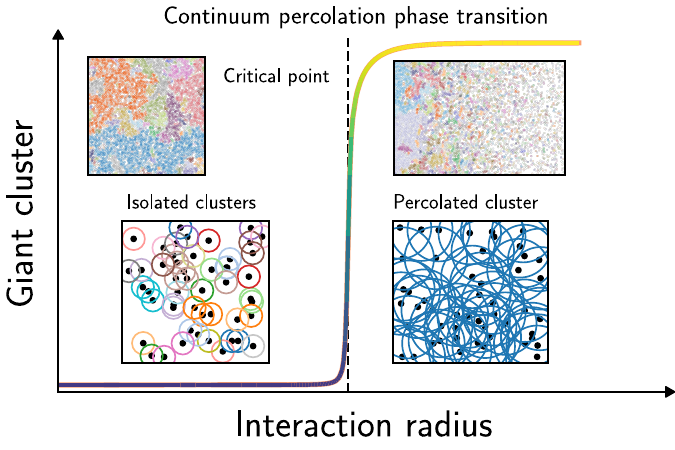}
\par\end{centering}
\caption{Sketch of the clustering process of a fixed number of points for different interaction distances. The upper insets illustrate the clusters at criticality for different homogeneous and inhomogeneous Poisson point processes (see \cite{Villegas2022} for further examples).\label{Sketch}}
\end{figure}

The most common way to characterize the aggregation properties of a fixed number of points, $N$, distributed in a continuous embedding space, relies on usual continuum percolation techniques \cite{gawlinski1981, Plotkin, Villegas2022}. Hence, based on some predefined distance $r$, two individual points will belong to the same cluster if their Euclidean distance is less than or equal to $r$ \cite{DBSCAN, DBSCAN2}. As illustrated in Figure \ref{Sketch}, a percolating cluster exists if a path can be drawn connecting all points with edges of length smaller than $r$. Let us remark that the \emph{mean nearest-neighbor} distance (MNN), defined as the nearest neighbor mean point-to-point Euclidean distance, makes it possible to study a point process independently of the area where it takes place and study it only depending on the system size \cite{Plotkin,Villegas2022}. Thus, we normalize the control parameter by the interparticle distance of the point process, i.e., defining $\hat r=r/\mbox{MNN}$ and producing a non-dimensional version of the distance parameter, which allows us to interpret cluster analysis in the language of statistical mechanics and percolation phase transitions.

The fraction of points belonging to the largest cluster, $P_{\infty}/N$, acts as the order parameter of the system, while the distance $\hat r$ acts as the control parameter, showing a percolation phase transition at some critical value $\hat r_{c}$. Additionally, it is possible to compute the susceptibility as $\chi(\hat r)=\nicefrac{\stackrel[S]{}{\sum}S^{2}P\left(S,\hat r\right)}{\stackrel[S]{}{\sum}SP\left(S,\hat r\right)}$, where the sum runs over all the possible sizes $S$ of the clusters (given a radius $\hat r$) in the system, being $P(S,\hat r)$ the cluster size probability distribution, and discarding $P_{\infty}$ if it exists (as usual). Let us highlight that $P(S,\hat r)$ is expected to follow a power law distribution only at the critical point, $P(S,\hat r_c)\sim S^{-\tau}$, where the Fisher exponent $\tau$ is linked to the intrinsic properties of the spatial point process \cite{Villegas2022} (e.g., in 2D, $\tau=2.05$ for isotropic percolation and $\tau=2.5\pm0.1$ for gradient percolation \cite{Villegas2022,Manna2022}).

\paragraph*{Barro Colorado Island (BCI).---} The BCI database comprises sufficiently high-resolution data with eight censuses (every five years from the 1980s) of more than $4\cdot 10^5$ trees and shrubs with diameter at breast height greater than $0.01\,$ m, belonging to about 300 species in 50-ha ($1000\times 500\,$m$^2$), and providing position and species for each plant \cite{condit2014}. From these data, we can resolve vegetation clusters of conspecific and heterospecific plants. Our primary goal is to provide a comprehensive overview, at the ecosystem level, of the intrinsic qualitative properties of the vegetation pattern by closely examining its aggregating properties.

Note that systems characterized by short-range correlation lengths (i.e., the absence of critical fluctuations) will be reflected in a standard percolation phase transition, where the properties of the underlying point process are determined by the Fisher exponent $\tau$ \cite{Villegas2022}. Let us emphasize that, e.g., critical systems lack a well-defined scale, so multiple broadly diverse scales are expected to make contributions of comparable importance, being microscopic, mesoscopic, and macroscopic scales all alike. Thus, a natural series of questions arises: Is it possible to extract information about characteristic spatial scales, in the case they emerge, at the ecosystem level? Do different functional scales live together in complex rainforests? Can we extract some information about the dynamical regime of these complex systems?

\begin{figure}[hbtp]
\begin{centering}
\includegraphics[width=1.0\columnwidth]{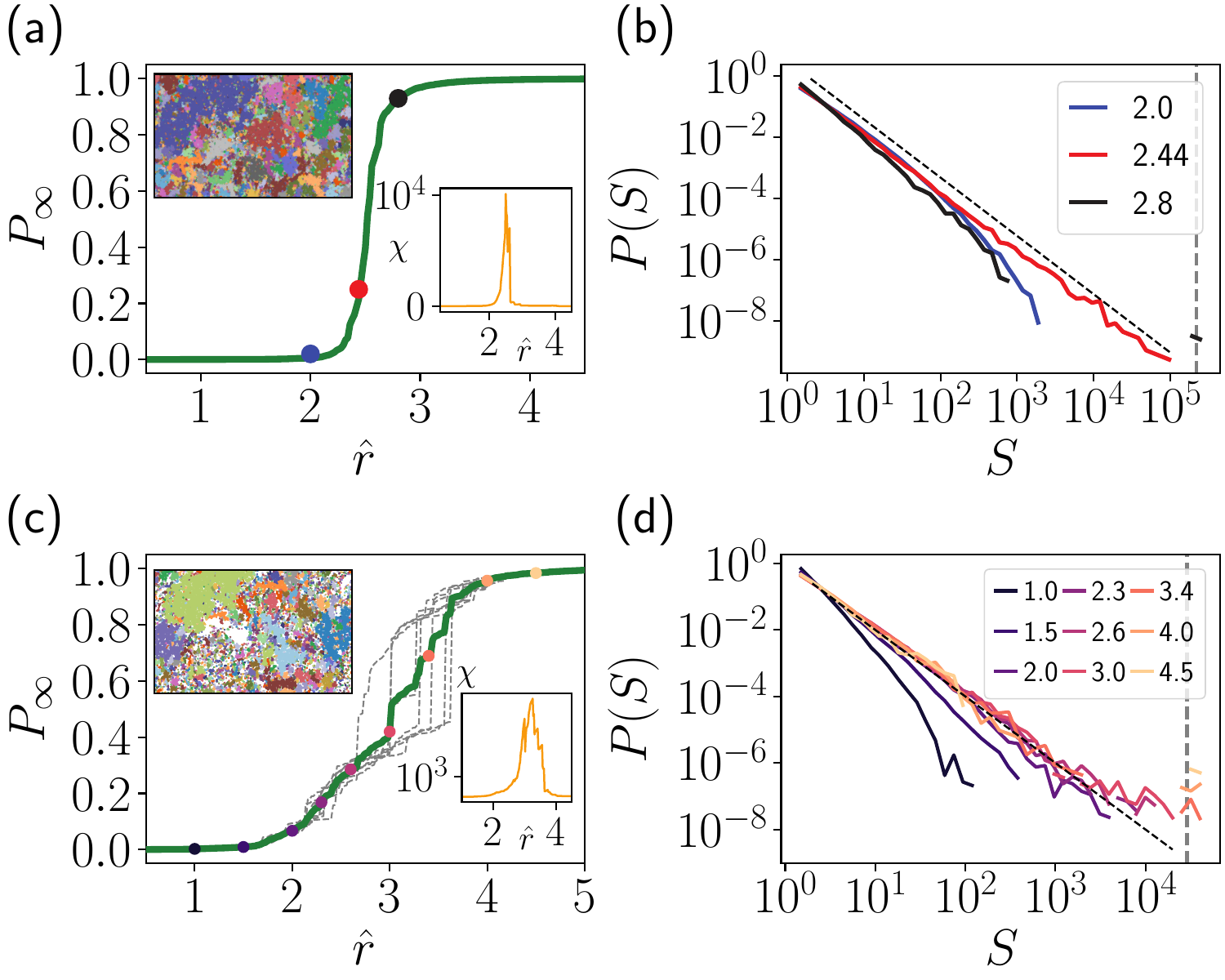}
\par\end{centering}
\caption{\textbf{Barro Colorado Island.} $P_{\infty}$ as a function of the distance parameter $\hat r$ for: \textbf{(a)} the community level and, \textbf{(c)} the most abundant species: $Hybanthus~prunifolius$. The solid green line shows the average phase transition over the eight censuses from 1985 to 2005, while the dashed gray lines represent individual censuses. The colored dots represent the position where the cluster size distributions are shown. Lower inset: Averaged system susceptibility. Upper inset: Spatial cluster distribution at $\chi_{max}$. $H.~prunifolius$ exhibits a non-trivial region with high susceptibility, ranging into $\hat r\in (2,4)$. \textbf{(b-d)} Cluster size distribution averaged over all censuses for different values of $\hat r$ (see legend). The community level shows a usual percolation transition with $\tau\simeq2.0\pm0.1$ (dashed lines are guides to the eye indicating this exponent.) A broad (critical) region with variable exponents exists for the most abundant species. The vertical lines represent the total system size (i.e., the number of individuals) for each specific case.\label{fig:PhTBCI}}
\end{figure}

We examine the aggregation properties of two important cases: the community- and single-species-levels. At the community level, we observe a percolation phase transition at approximately $\hat r_c\sim2.4\pm0.1$, as depicted in Fig.\ref{fig:PhTBCI}(a). At the critical point, different species aggregate, showing a power-law distribution of cluster sizes with an exponent of around $\tau\sim2.0\pm0.1$. This suggests the existence of a characteristic correlation length and reflects a short-range correlated distribution of points in a 2D space \cite{Villegas2022}. This result fully agrees with previous evidence of an explicit correlation length $\xi$ and analyses derived from the pair correlation function \cite{VillegasRS}. 

On the other hand, when we analyze the most abundant species (which has been shown to quantify the collective behavior of the entire system in agent-based models \cite{Villegas2021}), we observe an absolute lack of characteristic scales. In particular, there is a broad region where multiple resolution scales hierarchically percolate, as shown in Fig.\ref{fig:PhTBCI}(b). As a consequence, the distribution of clusters in the system is expected to follow an intrinsic power law for a wide range of $\hat r$ values. We will later analyze whether this region is only a finite-size effect or if we can explain it from a statistical physics perspective. We can consider the system inherently stationary for the specific analyzed period at the community- and most abundant species levels. Thus, we have performed the average over the eight censuses in Fig.\ref{fig:PhTBCI} for better quality data. However, analyzing species showing significant expansion or contraction is still of great interest. Additional results for these specific cases, and other species, can be found in Supp. Inf. \cite{SI}. We will later discuss the physical meaning and biological plausibility of this dynamic state.

\paragraph*{Spatial Explicit Neutral Model (SENM).---} As a matter of comparison, we analyze here the SENM \cite{Hubbell2001, Durrett1996, Pigolotti2018}, a non-equilibrium stochastic population model that captures the main features of ecological landscapes, allowing to generate non-trivial spatial patterns at the single-species level \cite{VillegasRS} (see also Supp. Inf. \cite{SI} for a thorough explanation of the SENM). Since we aim to observe robust collective phenomena, a SENM looks to be the suitable model to tackle our questions on real data, comparing them with synthetic point distributions. In the SENM, the magnitude of seed dispersal in all nodes is represented by the dispersal kernel $K$. The immigration rate $\nu$, i.e., the probability of having a new species at each timestep on each lattice point, is a parameter of species competition. Both parameters, $K$ and $\nu$, rule the phases and phase diagram of the system \cite{Villegas2021}. The rich phase diagram of SNEMs --characterized by the percolation properties of the most abundant species-- includes a short-range correlated region with random point patterns for high values of $K$, separated by a scale-invariant (or critical) region for short-range dispersal kernels and moderate values of $\nu$ and, finally, a bistable (monodominant) phase for low values of $K$ and $\nu$. In particular, the scale-invariant region exhibits finite-size scaling effects, with critical exponents belonging to the isotropic percolation universality class and ranging from 2D to mean-field behavior. For more details, we refer the reader to the original work \cite{Villegas2021}.

We analyze the two cases of particular interest for the most abundant species: (i) random and (ii) scale-invariant regions in the parameter space \cite{Villegas2021}. We therefore consider multiple independent realizations of the model in a square lattice of size $L=512$, exploiting the SENM duality with coalescing random walks \cite{Villegas2021,VillegasRS, Bramson1996,Pigolotti2018, Durrett1996, Holley1975}.

\begin{figure}[hbtp]
\begin{centering}
\includegraphics[width=1.0\columnwidth]{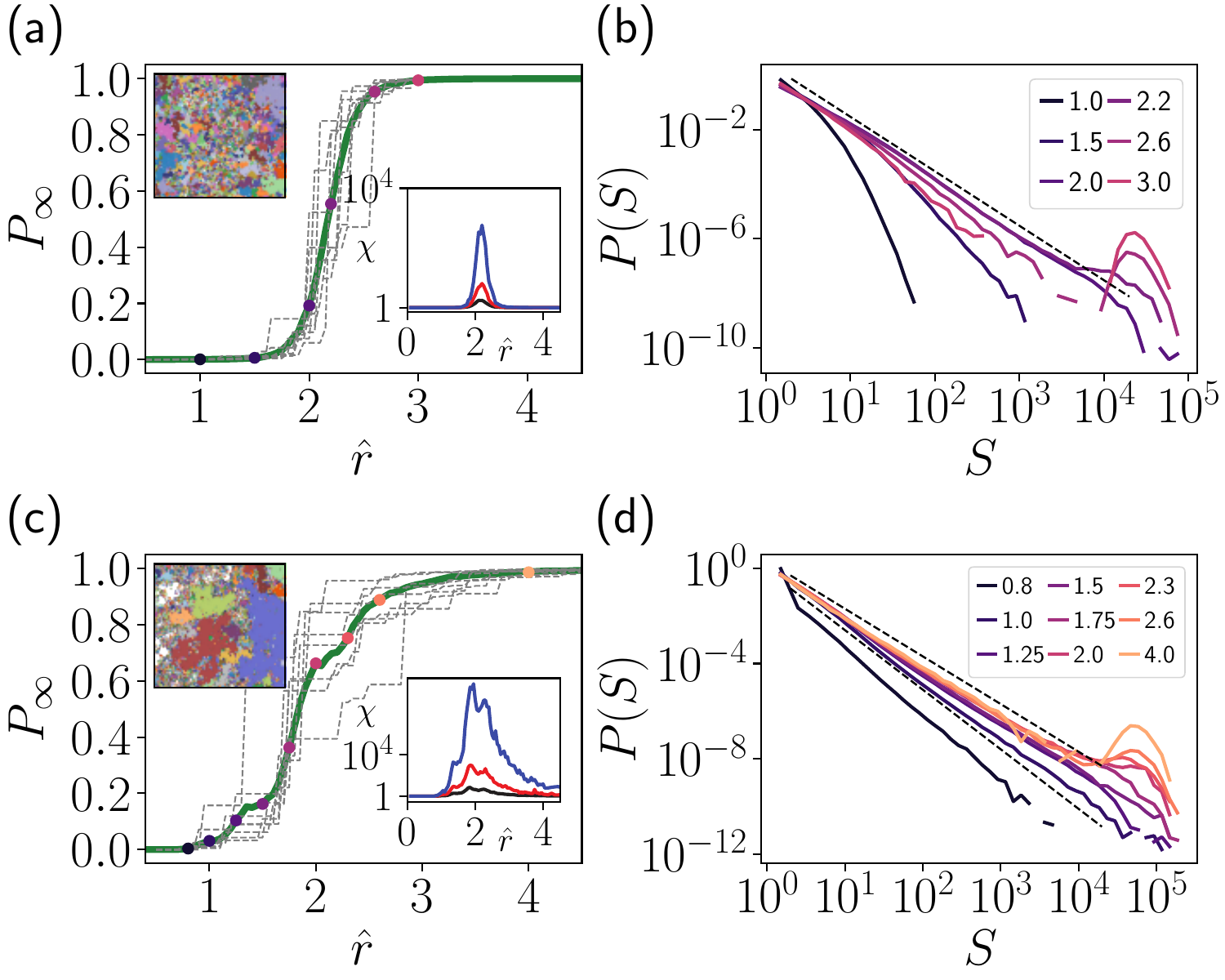}
\par\end{centering}
\caption{\textbf{SENM.} $P_{\infty}$ for the most abundant species as a function of the distance parameter, $\hat r$, for: \textbf{(a)} the random regime ($\nu=5\cdot10^{-6}, K=25, L=2^9$) and, \textbf{(c)} the scale-invariant regime ($\nu=5\cdot10^{-6}, K=5, L=2^9$). The solid green line shows the average phase transition over the $10^3$ independent realizations, while the dashed gray lines stand for the individual ones. The colored dots represent the position where the cluster size distributions are shown. Lower inset: Averaged system susceptibility over $10^3$ realizations for $L=2^9,2^{10}$ and $2^{11}$ (black, red, and blu lines, respectively). Note that $\chi$ diverges as the system size increases. For each lattice side, we change $\nu$ to maintain fixed $\nu L^2$, selecting it for $L=2^9$. Upper inset: Spatial cluster distribution at $\chi_{max}$. The scale-invariant regime exhibits a non-trivial extended region with divergent susceptibility. \textbf{(b)} and \textbf{(d)} $P(S)$ averaged over $10^3$ realizations for different values of $\hat r$ (see legend). The random regime exhibits a usual percolation phase transition, with $\tau\sim2.0\pm0.1$ (black dashed line), whereas the scale-invariant regime shows an intermediate region with variable exponents. Now, the dashed lines are guides for the eye for $\tau\sim2.0$ and $\tau\sim2.5$, respectively.\label{Neutral}}
\end{figure}

As can be seen, Figure \ref{Neutral}(a) shows the case of a large-scale dispersal kernel, leading to the emergence of short-range correlated (random) spatial patterns. We scrutinize the percolation transition of the most abundant species that displays a usual 2D phase transition at $\hat r\simeq2.39\pm0.01$ as expected for a Poisson point process \cite{Villegas2022}. Figure \ref{Neutral}(b) shows the Fisher exponent $\tau \simeq 2.0\pm0.1$ at the critical point, a sign of a single correlation length, $\xi$, for the global point process. However, the analysis of the scale-invariant phase in SENMs reveals a more complex situation. This theoretical scenario is analyzed in Fig.\ref{Neutral}(c), where the underlying percolation phase transition exhibits a lack of characteristic scales and heterogeneous long-range spatial correlations (resulting in an effective hierarchical phase transition). Moreover, the careful analysis of $P(S)$ for the most abundant species keeps track of a broad power-law regime with variable $\tau$, as shown in Fig. \ref{Neutral}(d). 

One crucial question is how to scale the system size to maintain the intrinsic properties of the SENM for analyzing the scale-invariant region. Recent studies have shown that the phase transition for the most abundant species results in a bonafide data collapse under the scaling relation $\nu L^2$ \cite{Villegas2021}. Thus, to ensure that the scaling of the most abundant species is consistent with the size of the system, we keep this specific value fixed while adjusting $\nu$ proportional to the increase in the size of the system. We refer to the Supp. Inf. \cite{SI} for a more detailed analysis based on different values of $\nu$.

In Figure \ref{Neutral}(a) and Figure \ref{Neutral}(c), the insets display the averaged susceptibility for systems of different sizes, with side $L$. Note that the scale-invariant region is characterized by divergent susceptibility when increasing the size of the system for a wide range of values of $\hat r$. From another angle, this fact confirms that this phase lacks a characteristic scale and qualitatively closely resembles the previous results obtained in the BCI, as illustrated in Figure \ref{fig:PhTBCI}(b). Most notably, it suggests that BCI may have evolved to operate at the edge of a second-order phase transition.

\begin{figure}[hbtp]
\begin{centering}
\includegraphics[width=1.0\columnwidth]{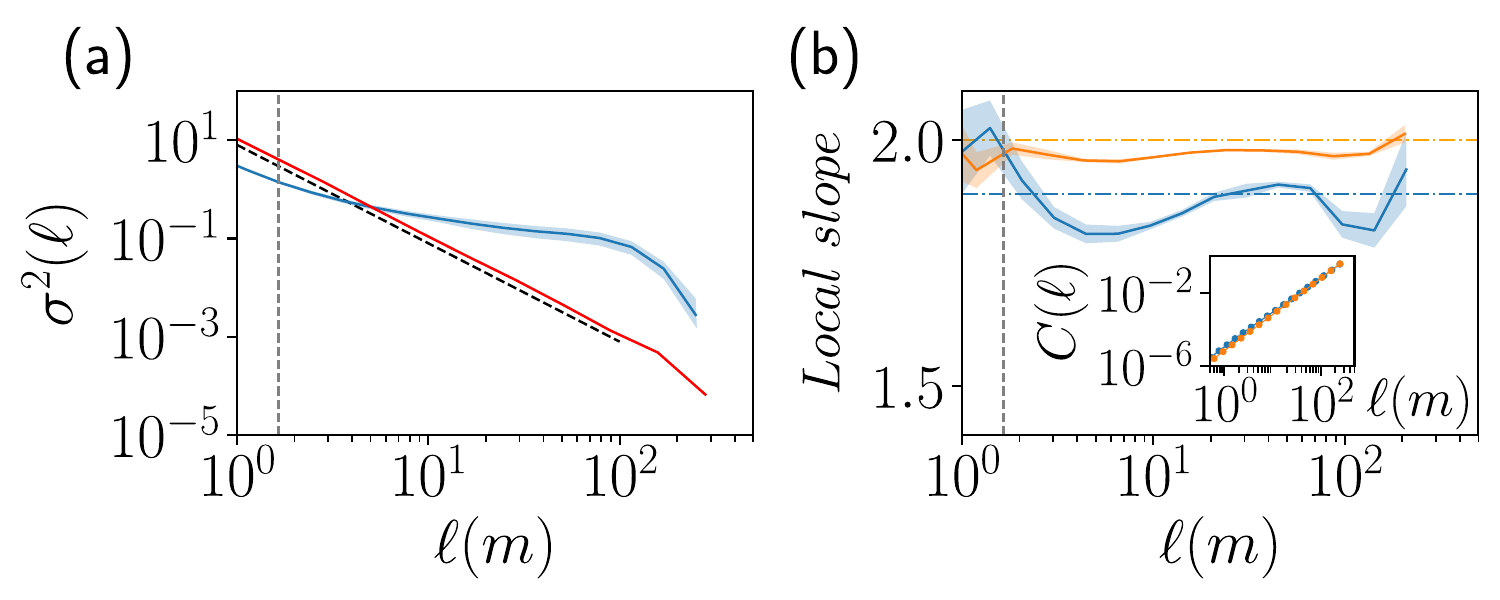}

\par\end{centering}
\caption{\textbf{Fractal nature of BCI. (a)} Mass variance, $\sigma^2(\ell)$, with different radius $\ell$ for $H.~prunifolius$ (blue line). Note the almost constant value, a sign of incipient criticality in the system. For the sake of comparison, the red line shows the Poisson case, with the same number of points and whose theoretical value scales as $\sigma^2(\ell)\propto\ell^{-2}$ (black dashed line). \textbf{(b)} Local slope for $C(\ell)$ (see inset) at the community level (orange line) and for $H.~prunifolius$ (blue line). We obtain $D=1.98\pm0.01$ for the community level and $D=1.86\pm0.04$ for $H.~prunifolius$. Gray dashed lines show the MNN distance for $H.~prunifolius$ in both figures. All curves have been averaged for the eigth available censuses (the shaded region shows a sigma confidence level). \label{Extra}}
\end{figure}

\paragraph*{Scale-invariant properties of BCI.---} To verify our hypothesis, we conducted additional tests to explore the potential critical behavior of the system. Specifically, we implemented different measures to capture the spatial scaling behavior of the point process \cite{GabrielliBook,Andrea2004}. These are based on the mean conditional density: that is, the number of point plants seen by another plant located at point $\mathbf{x}$ within a distance $\ell$ from it, $n_\ell(\mathbf{x})$. We thus compute this quantity adapting the Hanisch method to avoid system boundary effects \cite{hanisch1984,VillegasRS}: that is, excluding from the statistics of the neighbors those plants that are more distant from plant $i$ than the closest border of the system. On the one hand, we first study the \emph{conditional mass variance}, $\sigma^2(\ell)=\frac{\langle n^2_\ell(\mathbf{x})\rangle - \langle n_\ell(\mathbf{x})\rangle^2}{\langle n_\ell(\mathbf{x})\rangle^2}$, which in a (critical) fractal set must remain almost constant across all different scales (since relative fluctuations are approximately constant at all scales, due to the scale-invariance properties of the system). Note that, as it is shown in Fig. \ref{Extra}(a), the almost constant value of $\sigma^2(\ell)$ --the Poisson case scales as $\sigma^2(\ell)\propto\ell^{-2}$ in this particular case-- can be considered a hallmark of self-similar fluctuations.

On the other hand, we measure the correlation dimension, D, of the fractal set, being $C(\ell)=\frac{1}{N-1} \sum_{i=1}^{N_{c}(\ell)}\frac{n_\ell(\mathbf{x}_i)}{N_{c}(\ell)}$, where $N_{c}(\ell)$ is now the number of valid centers up to scale $\ell$. In particular, $C(\ell)$ scales as  $C(\ell) \propto \ell^D$. Fig.~\ref{Extra}(b) shows the local slopes for the community level and $H.~prunifolius$. We emphasize that the mean fractal dimension we found for $H.~prunifolius$, $D=1.86\pm0.04$ is fully compatible with the fractal dimension of 2D percolation clusters, $D=\frac{91}{48}$, at the (continuum) percolation threshold, i.e., with the \emph{isotropic percolation} universality class.

\begin{figure}[hbtp]
\begin{centering}
\includegraphics[width=1.0\columnwidth]{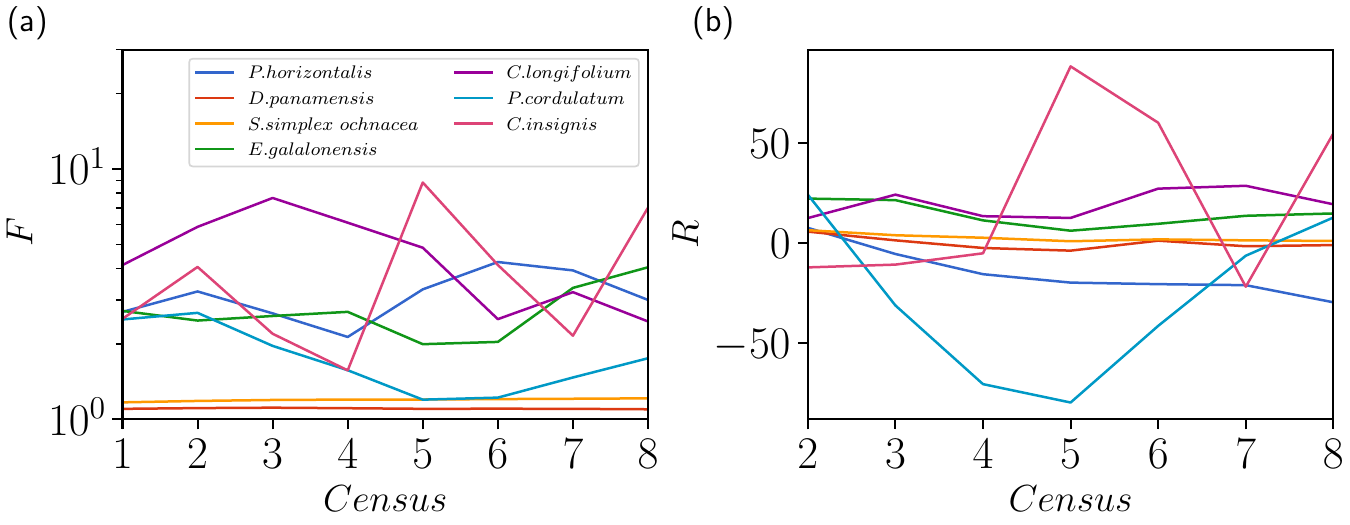}

\par\end{centering}
\caption{\textbf{Species fluctuations. (a)} Fluctuability index versus census number for different selected species in BCI (see legend). \textbf{(b)} Rate of change versus census number for the same selected species.  \label{Predic}}
 
\end{figure}

An essential consequence of the $\sigma^2(\ell)$ analysis involves characterizing fluctuation-driven species at short timescales. We point out that $\sigma^2(\ell_0)\approx1$ defines the scale, $\ell_0$, beyond which the average density becomes well defined \cite{GabrielliBook}. We then propose the \emph{fluctuability index}, $F=\frac{\ell_0}{MNN}$, which is a measure to assess to what extent the correlations are dominated by fluctuations: the higher the index, the higher the effect of fluctuations at the individual species dynamics (see Fig.\ref{Predic} (a)), potentially leading, at the single--species level, to catastrophic events or dramatic expansions over short time periods. To check the predictive relevance of this instant measure, we compare it with the rate of change for the $i^{th}$ species, defined as $R_i=\frac{N_i(t+1)-N_i(t)}{N_i(t)}$, where $N_i$ is the number of steams at every census for the $i^{th}$ species (see Fig.\ref{Predic} (b)). Note that $F$ can be a great predictor, that is, it facilitates anticipating and preventing catastrophic shifts at the single species level.

\paragraph*{Outlook.---} The hypothesis that some biological systems can extract significant functional benefits from operating in the vicinity of a critical point has recently thoroughly allowed the understanding of many real examples of natural phenomena and living matter \cite{MAM-RMP}. Setting close to criticality represents a simple strategy to balance the robustness (order) and flexibility (disorder) needed to derive functionality. Different works have recently shed light on multiple empirical examples spanning a wide range of biological systems taking advantage of operating close to critical dynamics: from neural dynamics \cite{BP2003}, to bacteria \cite{larkin2018}, macrophage dynamic \cite{nykter2008} or bird flocking \cite{NatComm}.

Here, we have shown that a saturated ecological environment such as BCI seems to be close to criticality, showing scale-free clusters of vegetation as evidenced by the spatial aggregating properties of the most abundant species. In particular, we have used existing approaches based on SENM --that have recently been demonstrated to exhibit critical percolating regions \cite{Villegas2021}-- to check the robustness of our results: we show the qualitative similarity between these two particular cases. This also strengthens recent theoretical frameworks that connect the
two-point correlation function to the distribution of species abundance and the relationship between species and area of a system at criticality \cite{AzaelePRX}. Furthermore, the most abundant species, $H. Prunifolius$ presents evidence of a dimension $D\simeq1.86\pm0.04$ --in complete agreement with previous analyses performed in low-canopy gaps \cite{sole1995}-- fully compatible with the expected value $D\sim1.89$ for percolating clusters in 2D.

Being aware that biological systems are finite, they cannot exhibit truly criticality in the narrow sense of statistical physics \cite{MAM-RMP}, but incipient scale-invariant features within the experimentally accessible ranges, as illustrated by mass fluctuations. In light of our findings, we hypothesize that the very particular distribution of vegetation patterns in saturated environments, such as the BCI one, can potentially benefit from being placed in the vicinity of a percolation critical point. Therefore, the most abundant species can potentially maintain high specific spatial correlations (i.e., clustering, maximizing the surface-to-bulk ratio) while, at the same time, they can explore a large proportion of the available space (as expected for large-scale dispersal kernels leading to short-range correlated distributions of points \cite{Villegas2021}). This maximizes the interaction with other species for nutrients, etc.

We have examined a specific model that considers a fixed dispersal kernel and only allows for the local extinction of particular species. However, our results confirm that local dispersal kernels can be a fundamental ingredient when considering spatial neutral dynamics \cite{Villegas2021}. Also, SENM models usually describe homogeneous saturated ecosystems, where population extinction was inevitable even with different dispersal kernels in cases of inbreeding depression or extreme environmental variability. Therefore, it is essential to develop more realistic perspectives integrating the intricate multiscale spatial structure and temporal variability of ecological ecosystems to better understand their overall stability \cite{Metz}. On the one hand, further work is still needed in incorporating, for instance, bet-hedging strategies, which are known to facilitate the viability of populations for more extended periods of time \cite{Hidalgo}, or including possible global extinction events (that is, the possibility to reach an absorbing state after which the vegetation cannot recover \cite{NicolettiFires, Paula}).

On the other hand, it is also essential to integrate principles of metapopulation dynamics, a core concept in theoretical ecology, which is already known to be crucial in mitigating the effect of local extinctions as it promotes migration from other sources  \cite{Hanski2000, HanskiBook}. For example, the pioneering work by Hanski and Ovaskainen demonstrated how survivability can depend on the specific fragmentation of a landscape from phenomenological metapopulation models \cite{Hanski2000}. In this respect, recent results have evidenced the relevance of modular structures and interconnected hubs to display an optimal metapopulation capacity, showing that isolation can harm survivability. At the same time, sparsity, in general, can drastically reduce species' persistence \cite{Nicoletti}.

Let us pinpoint that we have not explored potential self-organization mechanisms to operate in the vicinity of such a critical point without resorting to parameter fine-tuning. Dispersal mechanisms, niche and adaptive effects, and the possibility of including diverse interacting patches or ecosystems will be analyzed in future work.
Also, our results serve as the basis for monitoring highly fluctuating species and open the door to further analysis, which can help predict and eventually prevent catastrophic shifts in these ecosystems by providing macroscopic evidence that criticality can be a symptom of a healthy and resilient ecosystem.
\begin{acknowledgments}
  P.V. acknowledges financial support from the Spanish 'Ministerio de Ciencia e Innovaci\'on' and the 'Agencia Estatal de Investigaci\'on (AEI)' under Project Ref. PID2020-113681GB-I00. G.C. acknowledges the EU project 'HumanE-AI-Net,' no. 952026. We also thank M.A. Mu\~noz, A. Maritan, and S. Suweis for extremely valuable discussions and/or suggestions on earlier versions of the manuscript.
\end{acknowledgments}

\def\url#1{}
%

\clearpage


\section*{Supplementary material}

# Supplementary information: Evidence of scale-free clusters of vegetation in tropical rainforests

Pablo Villegas,[1,2,*] Tommaso Gili,[3] Guido Caldarelli,[4,5,6,7] and Andrea Gabrielli[1,8,6]

[1]*'Enrico Fermi' Research Center (CREF), Via Panisperna 89A, 00184 - Rome, Italy.*
[2]*Instituto Carlos I de Física Teórica y Computacional,
Universidad de Granada, E-18071, Granada, Spain.*
[3]*Networks Unit, IMT Scuola Alti Studi Lucca, Piazza San Francesco 15, 55100 - Lucca, Italy.*
[4]*DMSN, Ca' Foscari University of Venice, Via Torino 155, 30172 - Venice, Italy.*
[5]*European Centre for Living Technology (ECLT), Dorsoduro 3911, 30123 - Venice, Italy.*
[6]*Institute for Complex Systems (ISC), CNR, UoS Sapienza, Piazzale Aldo Moro 2, 00185 - Rome, Italy.*
[7]*London Institute for Mathematical Sciences (LIMS), W1K2XF London, United Kingdom.*
[8]*Dipartimento di Ingegneria Civile, Informatica e delle Tecnologie Aeronautiche,
Università degli Studi âRoma Treâ, Via Vito Volterra 62, 00146 - Rome, Italy.*

## CONTENTS

---

[*] Corresponding author: pablo.villegas@cref.it

## I. THE SPATIAL EXPLICIT NEUTRAL MODEL (SENM)

The dynamics of the neutral model in a spatially explicit 2D environment can be explained through the following steps. Imagine a square lattice of size N ($L^2$). First, a random tree dies and is replaced by either of the following events:

- Dispersal effect: A new tree is selected from the neighborhood with a probability $1 - \nu$. The neighborhood is defined as a square dispersal kernel of size K, centered on the gap left by the dead tree.

- Speciation or immigration event: A new species is introduced with probability $\nu$.

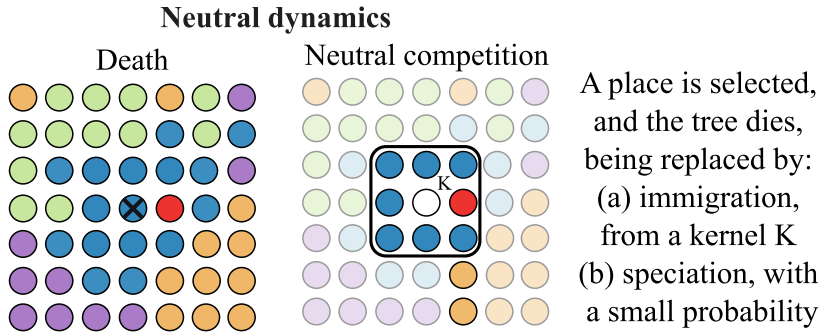

FIG. 1. Sketch of the neutral dynamics.

***Dual representation of the SENM*** SENM simulations use a clever method of merging random walkers to achieve duality [1–3]. This approach is widely employed for its remarkable efficiency in producing high-speed simulations of independent realizations without interfering with boundary effects found in periodic boundary conditions (typically used in the model's forward dynamics). However, this strategy is limited to the examination of the long-term static properties of the model, making it an ideal tool for studying the ergodic properties of infinite systems [3]. Hence, this allows for generating many sample patterns at the (non-equilibrium) stationary state for exploring the whole statistical ensemble.

The dual dynamics of the SENM works as follows (see also [4, 5] for an accurate description of the process):

- The process of tracing the ancestry of a species involves placing a random walker on each lattice site and moving backward in time.

- With probability $1-\nu$, a randomly chosen walker is moved –at each discrete (backward) time step– to a different site selected from a squared dispersal kernel of length $K$ (pay particular attention to the possibility of choosing a site outside the sampled domain, since we only observe a portion of the infinite lattice). If two walkers end up on the same site, they merge, removing one and tracing the coalescing partner.

- Alternatively, a walker may be killed with a probability of $\nu$, representing a speciation/migration event in the forward dynamics.

- The simulation ends when only one walker remains, and as walkers decrease with each event, the simulation time speeds up.

The complete history of coalescing events, also called the tree of coalescences, allows us to track a complete genealogical tree of a species back to its origin at the speciation event. To further explain this, when all walkers in a dual simulation coalesce or are annihilated, the species are assigned to the starting point of each walker, creating a potential configuration. This configuration can be viewed as a snapshot of the set of possible states within the SENM [4, 5].

3

## A. Percolation transitions in the SENM

Computational analyses in 2D systems have revealed a rich phenomenology in the SENM (as graphically illustrated, for instance, in Fig. 2). We refer to [6] for an extended discussion on the issue. The SENM exhibits three different types of characteristic regimes (i.e., spatial point patterns), in a nutshell: a region of bistability (which shows monodominance or uncorrelated point patterns among a small number of species), a random region (for large dispersal kernels), and a scale-invariant one (characterized by scale-free distributed clusters of vegetation, with exponents compatible with percolation transitions [6].)

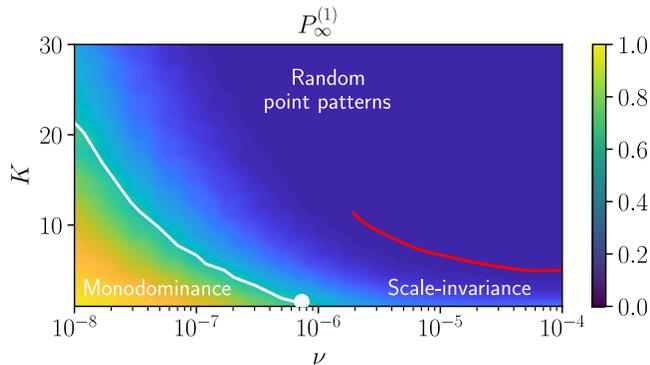

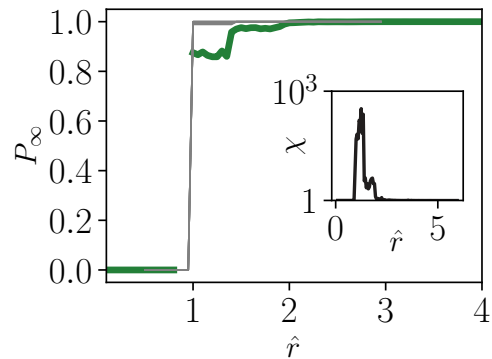

FIG. 2. Phase diagram for the most abundant species. Colors represent the intensity of the percolation strength for the most abundant species, $P_\infty^{(1)}$. The white line (involving bistability, which vanishes at the white point) divides monodominant states, where more than 50% of the tree canopy comprises a single tree species, to coexistence between a small number of species. The red line shows the estimated limit of the scale-invariant regime. Non-homogeneous random point patterns emerge for large dispersal kernels (bluish regions). Simulations have been averaged over $10^3$ runs on a lattice of size $N = 512^2$.

FIG. 3. $P_\infty$ for the most abundant species, in a lattice of square size $L = 512$, as a function of the distance parameter, $\hat{r}$. Note the discontinuous phase transition, which can detect the bistable nature of the selected point in the phase diagram. Parameters: $N = 512^2$, $K = 5$, $\nu = 10^{-8}$. The green curve has been averaged over $10^3$ independent realizations.

We gain insights into the distinct emergent point patterns of the SENM by precisely examining this phase diagram. First, Figure 3 shows a specific example of a spatial pattern belonging to the monodominant (bistable) regime. We highlight the discontinuous transition at a particular value $\hat{r}_c$. To confirm the broad phase transition with divergent susceptibility across diverse $\hat{r}$ values, we examined the percolation phase transition (as elucidated in the main text) at different locations within the parameter space, as shown in Figure 4.

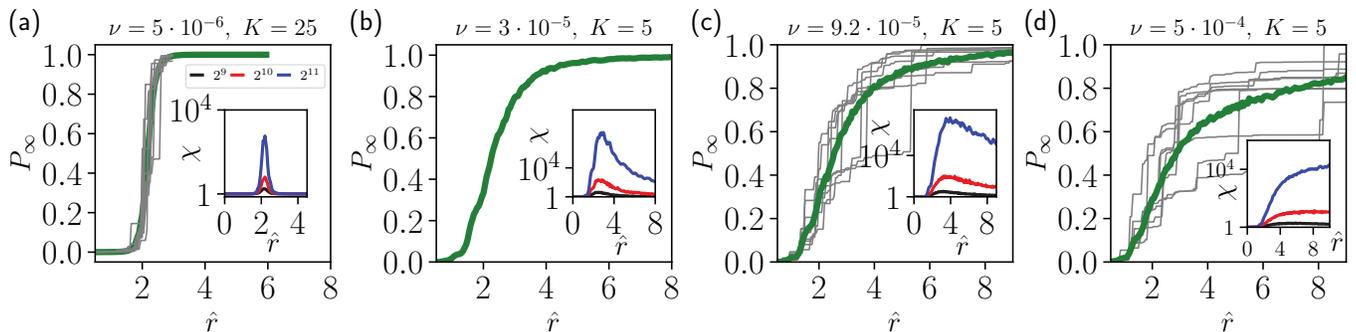

FIG. 4. $P_\infty$ for the most abundant species, in a lattice of square size $L = 512$, as a function of the distance parameter, $\hat{r}$, for: (a) the random regime and, (b)-(d) the scale-invariant regime. The green line shows the average phase transition over $10^3$ independent realizations, while grey lines represent individual ones. Inset: Averaged system susceptibility over $10^3$ realizations. We change $\nu$ to maintain fixed $\nu L^2$, selecting it for $L = 2^9$. Note that the scale-invariant regime exhibits a non-trivial extended region with divergent susceptibility.

4

## II. DISTRIBUTION OF CLUSTER SIZES FOR DIFFERENT SPECIES IN BCI

### A. Most abundant species

We present further analyses based on some of the most abundant species in Barro Colorado Island. As shown in Figures 5 and 6, the reported results for *H prunifolius* are robust for some of the most abundant species in BCI. In particular, both *Tetragastris panamensis* and *Trichilia tuberculata* are part of the ten most abundant species group, showing an intermediate state with scale-free clusters of vegetation.

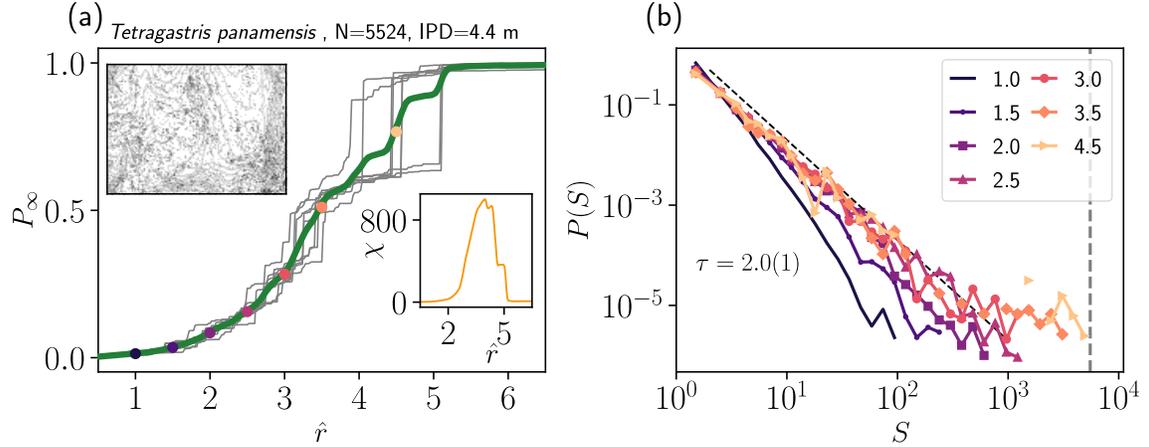

FIG. 5. **_Tetragastris panamensis._** (a) $P_\infty$ as a function of the distance parameter $\hat{r}$. The green line shows the average phase transition over the eight censuses from 1985 to 2005, while the grey lines represent individual censuses. Colored dots stand for the position where cluster size distributions are shown. Lower inset: Averaged system susceptibility. A non-trivial extended region with high susceptibility, ranging into $\hat{r} \in (2, 4)$, can be appreciated. Upper inset: Spatial distribution for Census 8. (b) Cluster size distribution averaged over all censuses for different values of $\hat{r}$ (see legend). An intermediate (critical) region exists with variable exponents. Dashed lines are guides to the eye representing $\tau = 2.0$.

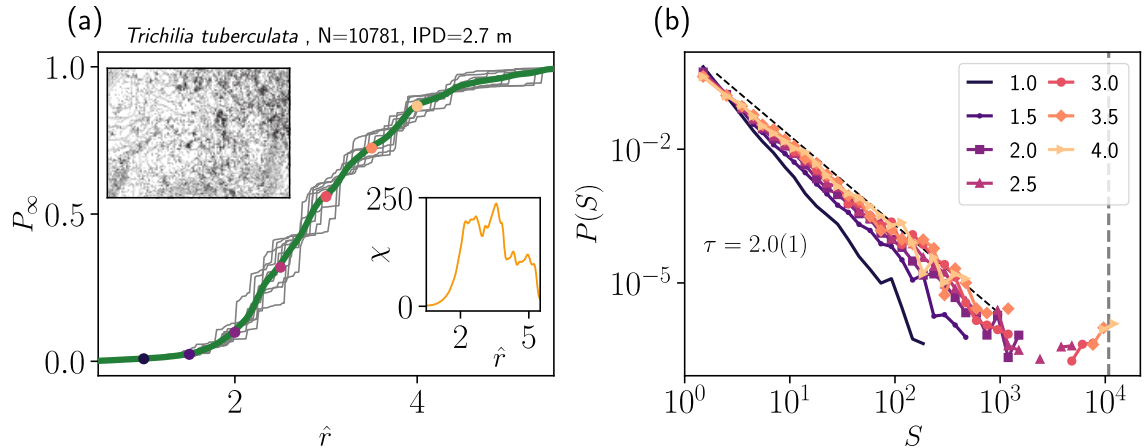

FIG. 6. **_Trichilia tuberculata._** (a) $P_\infty$ as a function of the distance parameter $\hat{r}$. The green line shows the average phase transition over the eight censuses from 1985 to 2005, while the grey lines represent individual censuses. Colored dots stand for the position where cluster size distributions are shown. Lower inset: Averaged system susceptibility. A non-trivial extended region with high susceptibility, ranging into $\hat{r} \in (2, 5)$, can be appreciated. Upper inset: Spatial distribution for Census 8. (b) Cluster size distribution averaged over all censuses for different values of $\hat{r}$ (see legend). An intermediate (critical) region exists with variable exponents. Dashed lines are guides to the eye representing $\tau = 2.0$.

## B. Expanding species

We present results for two of the most notable cases of expanding species in BCI. In particular, we select *Eugenia galanonensis* (it has risen from $N = 964$ stems in Census 1 to $N = 2400$ stems in Census 8) and *Calophyllum longifolium* (it has risen from $N = 647$ stems in Census 1 to $N = 2237$ stems in Census 8).

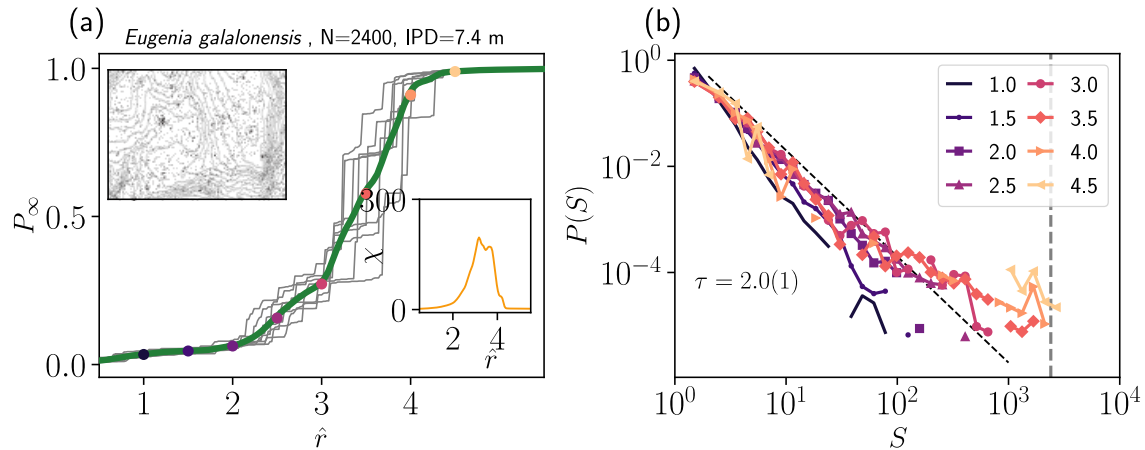

FIG. 7. ***Eugenia galanonensis.*** **(a)** $P_\infty$ as a function of the distance parameter $\hat{r}$. The green line shows the average phase transition over the eight censuses from 1985 to 2005, while the grey lines represent individual censuses. Colored dots stand for the position where cluster size distributions are shown. Lower inset: Averaged system susceptibility. A non-trivial extended region with high susceptibility, ranging into $\hat{r} \in (2,5)$, can be appreciated. Upper inset: Spatial distribution for Census 8. **(b)** Cluster size distribution averaged over all censuses for different values of $\hat{r}$ (see legend). An intermediate (critical) region exists with variable exponents. Dashed lines are guides to the eye representing $\tau = 2.0$.

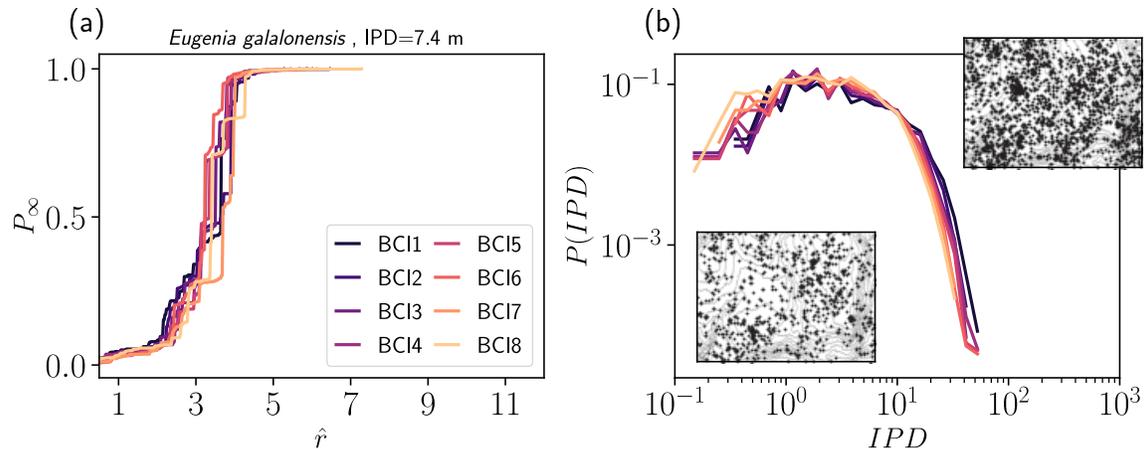

FIG. 8. ***Eugenia galanonensis.*** **(a)** $P_\infty$ as a function of the distance parameter $\hat{r}$ for the eight censuses from 1985 to 2005 (see legend). **(b)** Mean-nearest neighbor distance probability distribution for the different censuses. Insets: Species spatial distribution for Census 1 (lower inset) and Census 8 (upper inset), respectively.

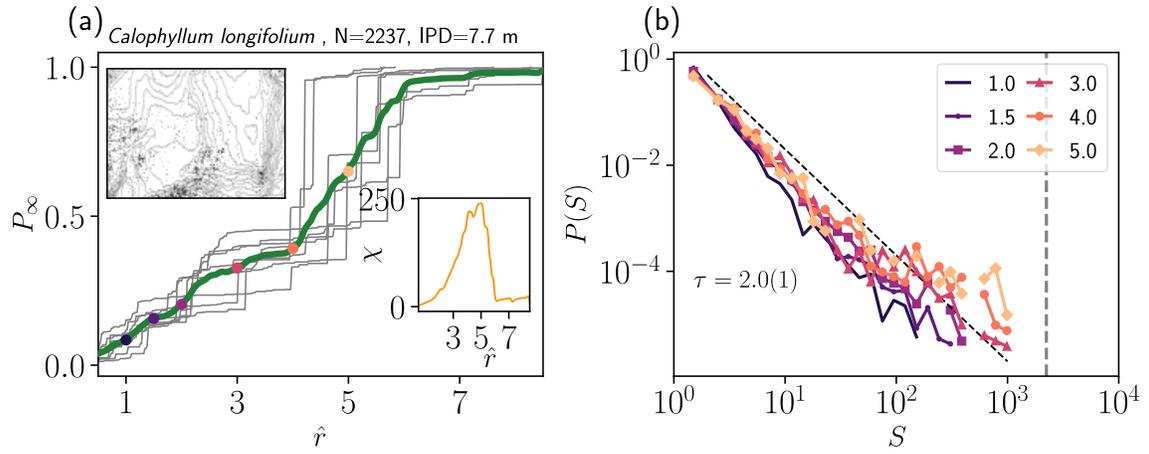

FIG. 9. ***Calophyllum longifolium.*** **(a)** $P_\infty$ as a function of the distance parameter $\hat{r}$. The green line shows the average phase transition over the eight censuses from 1985 to 2005, while the grey lines represent individual censuses. Colored dots stand for the position where cluster size distributions are shown. Lower inset: Averaged system susceptibility. A non-trivial extended region with high susceptibility, ranging into $\hat{r} \in (2, 5)$, can be appreciated. Upper inset: Spatial distribution for Census 8. **(b)** Cluster size distribution averaged over all censuses for different values of $\hat{r}$ (see legend). An intermediate (critical) region exists with variable exponents. Dashed lines are guides to the eye representing $\tau = 2.0$.

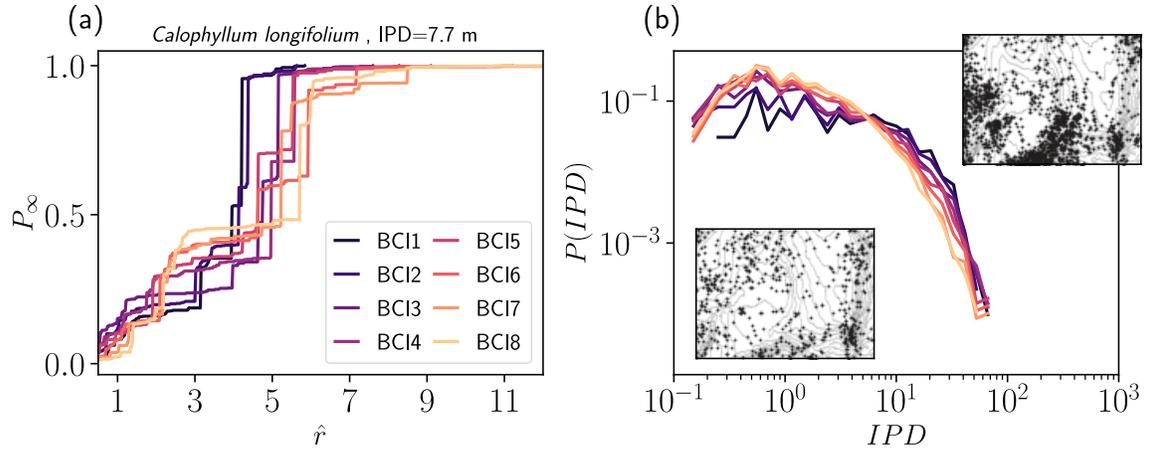

FIG. 10. ***Eugenia galanonensis.*** **(a)** $P_\infty$ as a function of the distance parameter $\hat{r}$ for the eight censuses from 1985 to 2005 (see legend). **(b)** Mean-nearest neighbor distance probability distribution for the different censuses. Insets: Species spatial distribution for Census 1 (lower inset) and Census 8 (upper inset), respectively.

### C. Contracting species

We present results for two of the most notable cases of contracting species in BCI. In particular, we select *Piper cordulatum* (it has risen from $N = 3142$ stems in Census 1 to $N = 98$ stems in Census 8) and *Psychotria horizontalis* (it has risen from $N = 6159$ stems in Census 1 to $N = 1856$ stems in Census 8).

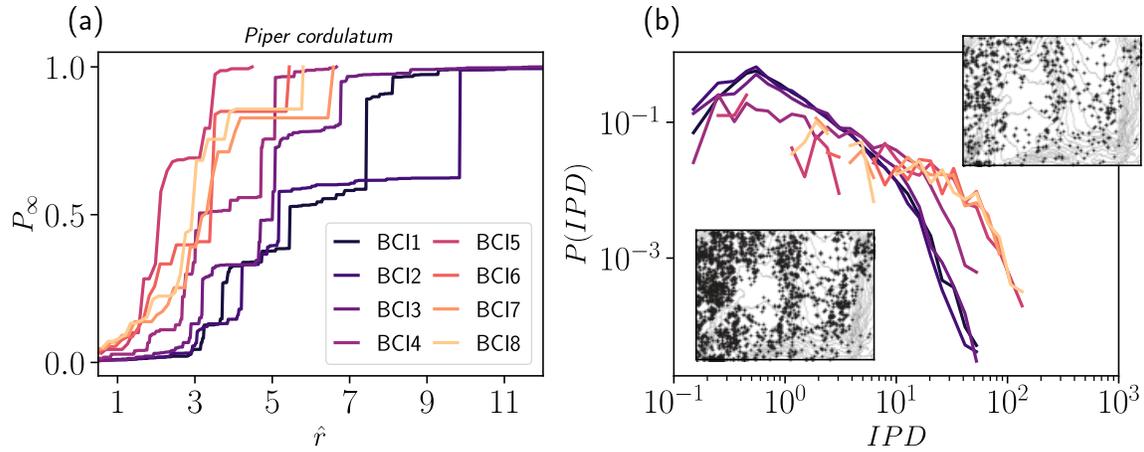

FIG. 11. ***Piper cordulatum.*** **(a)** $P_\infty$ as a function of the distance parameter $\hat{r}$ for the eight censuses from 1985 to 2005 (see legend). **(b)** Mean-nearest neighbor distance probability distribution for the different censuses. Insets: Species spatial distribution for Census 3 (lower inset) and Census 4 (upper inset), respectively.

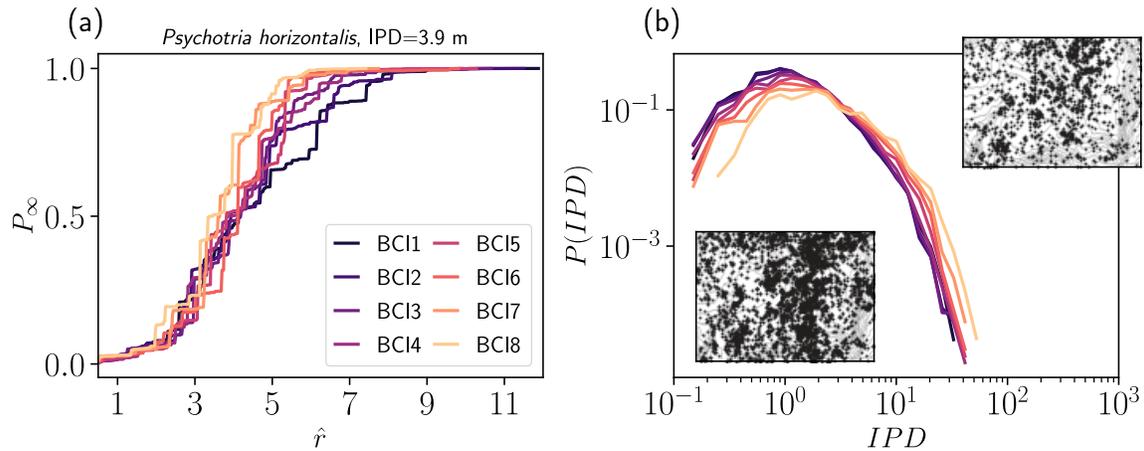

FIG. 12. ***Psychotria horizontalis.*** **(a)** $P_\infty$ as a function of the distance parameter $\hat{r}$ for the eight censuses from 1985 to 2005 (see legend). **(b)** Mean-nearest neighbor distance probability distribution for the different censuses. Insets: Species spatial distribution for Census 3 (lower inset) and Census 4 (upper inset), respectively.


\end{document}